\documentclass{revtex4}
\usepackage{amsmath,amsthm}
\usepackage{rotating}
\usepackage{blkarray}
\usepackage{booktabs}
\usepackage{mathtools}
\usepackage{multirow}
\usepackage{adjustbox}

\usepackage{graphicx}

\usepackage{amsfonts}
\newtheorem{theorem}{Theorem}[section]
\newtheorem{lemma}[theorem]{Lemma}
\newtheorem{proposition}[theorem]{Proposition}
\newtheorem{cor}[theorem]{Corollary}

\theoremstyle{remark}
\newtheorem{remark}[theorem]{Remark}

\theoremstyle{definition}
\newtheorem{definition}[theorem]{Definition}

\theoremstyle{example}
\newtheorem{example}[theorem]{Example}

\theoremstyle{notation}

\newcommand{\bra}[1]{\langle#1|}
\newcommand{\ket}[1]{|#1\rangle}

\begin{document}

\title{Partial order and  topology of Hermitian matrices and quantum Choquet integrals for density matrices with given expectation values}            
\author{A. Vourdas}
\affiliation{Department of Computer Science,\\
University of Bradford, \\
Bradford BD7 1DP, United Kingdom\\a.vourdas@bradford.ac.uk}

\begin{abstract}
The set $M$ of $d\times d$ Hermitian matrices (observables) is studied as a partially ordered set with the  L\"{o}wner partial order.
Upper and lower sets in it, define the concept of cumulativeness (used mainly with scalar quantities) in the context of Hermitian matrices.
Partial order and topology are intimately related to each other and the  set $M$ of Hermitian matrices is also studied as a topological space,
 where open and closed sets are the upper and lower sets. It is shown that the set $M$ of Hermitian matrices is a $T_0$ topological space, and its subset ${\mathfrak D}$ of density matrices is Hausdorff totally disconnected topological space.
These ideas are a prerequisite for studying quantum Choquet integrals with Hermitian matrices (as opposed to classical Choquet integrals with scalar quantities). 
Capacities (non-additive probabilities), cumulative quantities that involve Hermitian matrices, and M\"obius transforms that remove the overlaps between non-commuting observables, are used in quantum Choquet integrals.
An application of the formalism is to find a density matrix, with given expectation values with respect to $n$ (non-commuting) observables.
Examples of calculations of such a density matrix (with quantified errors in its expectation values), are presented.

\end{abstract}
\maketitle

\section{Introduction}

Capacities (non-additive probabilities), cumulative functions and Choquet integrals \cite{C}
have been used extensively with scalar quantities in Artificial Intelligence, Operational Research and Mathematical Economics (e.g.,\cite{D1,D2,D3,D4,D5,D6,D7}).
These techniques are very important in recent developments in Artificial Intelligence and Big Data (e.g. \cite{A1,A2,A3,A4,A5}).
They are used in situations where the various alternatives {\bf are not truly independent (in a quantum context this corresponds to non-orthonormal bases and non-commuting observables)}. When the various alternatives overlap (in a quantum context the observables do not commute) we need a mechanism of removing the overlaps, and this is the  
M\"obius transform that plays an important role within Choquet integrals.
In a quantum context Choquet integrals have been studied in \cite{Q1,Q2,Q3,Q4}.

In this paper we study quantum Choquet integrals that involve integration of Hermitian matrices (observables). 
A prerequisite for capacities, cumulative functions and Choquet integrals is an ordering of the function (e.g., Eq(\ref{order}) in a 
classical context) which is trivial in the case of scalar quantities, but needs clear definition in the case of matrices.
We use the L\"{o}wner partial order with the set $M$ of $d\times d$ Hermitian matrices\cite{L1,L2}, and study its properties.
In particular we introduce upper and lower sets which extend the physical concept cumulativeness.
Cumulativeness with scalar quantities in totally ordered sets, has been used extensively in probability theory, Mathematical Economics, Finance , etc. 
In this paper we extend it to the partially ordered set of Hermitian matrices using upper and lower sets.

We also discuss the partial ordering of Hermitian matrices using the language of topology.
There are deep links between partial order theory and topology (e.g., \cite{A}).
Here we study the set $M$ of Hermitian matrices as a topological space (the upper and lower sets become open and closed sets, correspondingly).
We show that $M$ is a $T_0$ topological space, and that its subset ${\mathfrak D}$ of density matrices is Hausdorff totally disconnected topological space.
We also explain that  topological concepts in a Physics context, describe logical relations between physical quantities.

The partial order of Hermitian matrices is a prerequisite for defining quantum Choquet integrals.
We define capacities (non-additive probabilities) in a quantum context, and quantum Choquet integrals.
They involve M\"obius transforms, which in the present context remove the overlaps between non-commuting observables.

Quantum Choquet integrals are used for the following physical problem.
We consider $n$ positive definite observables $\theta(\{1\}),...,\theta(\{n\})$ (which in general do not commute with each other) and the 
corresponding expectation values $\langle \theta(r)\rangle$ with some unknown density matrix $R$.
We take $d\le n<d^2-1$, in which case the $n$ expectation values are  incomplete information about $R$.
We show that  quantum Choquet integrals lead to a family of {\bf positive semi-definite} matrices 
$R_1$ which are an approximation to the density matrix $R$, in the sense that the expectation values of $R_1$ are close to $\langle \theta(r)\rangle$ with errors which are quantified.

This problem can be viewed as a generalisation of quantum tomography (e.g., \cite{tomo}), which uses very specific observables for the calculation of the density matrix.
In the present paper the observables are arbitrary, and the technical details are very different.
Apart from the practical value of this problem, there is theoretical merit in the sense that we approach quantum non-commutativity, with capacities and quantum Choquet integrals.
We also show that in the special case of orthonormal bases, the capacities become Kolmogorov (additive) probabilities and the Choquet integral gives standard results.

In section 2 we define several quantities which are needed later.
In section 3 we present the L\"{o}wner partial order and some of its properties.
We also show that many physical quantities in widely used expansions are non-comparable with respect to this partial order, and this differentiates these techniques from our approach.
In section 4 we discuss upper and lower sets, which define cumulativeness in the set of observables (Hermitian matrices).
In section 5 we use the topological language (which is intimately related to the partial order language) to describe these concepts.
Sections 3,4,5 are  interesting in their own right, but here they are presented as prerequisites for quantum Choquet integrals.

In section 6 we apply these ideas to quantum Choquet integrals.
We first discuss briefly capacities and Choquet integrals in a classical context (section \ref{sec13}).
We then introduce in section \ref{sec14} capacities in a quantum context (the partial order among the Hermitian matrices is essential here). 
In section \ref{sec15} we introduce the M\"obius transform of capacities, which quantify their deviations from additivity.
In section \ref{sec16} we present the physical problem that we are going to solve with quantum Choquet integrals, which is to find a density matrix from its expectation values with several observables.
We then present quantum Choquet integrals (section \ref{sec17}).

In section 7 we present examples.
We conclude in section 8 with a discussion of our results and open problems.

\section{Preliminaries}
\subsection{Notation}

We label the eigenvalues $e_r(\theta)$ of a $d\times d$ Hermitian matrix $\theta$ in ascending order:
\begin{eqnarray}\label{11}
e_1(\theta)\le e_2(\theta)\le...\le e_d(\theta).
\end{eqnarray} 
The interval ${\cal I}(\theta)$ of all the values of the quadratic form $\sum _r x_r^*\theta _{rs}x_s$, for all normalised vectors $x$ with $||x||=1$,
is known to be ${\cal I}(\theta)=[e_1(\theta),e_d(\theta)]$ \cite{M} .
Equivalently, ${\cal I}(\theta)$ is the set of all values of $\sum _r x_r^*(U\theta U^\dagger)_{rs}x_s$, for all unitary transformations $U$, and a fixed vector $x$ with $||x||=1$.

A Hermitian $d\times d$ matrix $\theta$ is positive semi-definite (denoted as $\theta\succ {\bf 0}$) if 
all its eigenvalues are non-negative, or equivalently if for all $d$-dimensional vectors (columns) $x$, we get $x^\dagger \theta x\ge 0$.

We use the following notation:
\begin{itemize}
\item
$M$ is the set of $d\times d$ Hermitian matrices. The sum of two matrices in $M$, belongs to $M$. The product of two matrices in $M$, belongs to $M$ only if the two matrices commute.
\item
$M_P$ is the subset of $M$, that contains positive semi-definite matrices (matrices with $e_1(\theta)\ge 0$).

\item
${\mathfrak D}$ is the subset of $M_P$ that contains all density matrices. They have trace one, and they are denoted with upper case letters.

\item
$M_N$ the subset of $M$, that contains negative semi-definite matrices (matrices with $e_d(\theta)\le 0$).
\item
$M_{PN}=M\setminus (M_P\cup M_N)$ is the subset of $M$, with non-definite matrices (matrices with $e_1(\theta)< 0<e_d(\theta)$).
\end{itemize}
The sets $M, M_P, M_N, {\mathfrak D}$ are convex.

\subsection{Quantum systems with variables in ${\mathbb Z}_d$ and coherent states}

We consider a quantum system with variables in ${\mathbb Z}_d$ (the ring of integers modulo $d$), described with  a $d$-dimensional Hilbert space $H(d)$ (e.g., \cite{VOU}). 
There are known differences between the cases of odd and even dimension $d$, and here we consider the case of odd $d$.
For example, the $2^{-1}=\frac{d+1}{2}$ (used below in Eq.(\ref{AA1}) exists in ${\mathbb Z}_d$ because $d$ is an odd integer.
Also the Wigner function (which is not used here) is more complicated (`doubled') in the case of even $d$ (e.g., \cite{zak}).

$|X;r\rangle$ where $r\in {\mathbb Z}_d$, is an orthonormal basis which we call position basis (the $X$ in this notation is not a variable, but it simply indicates position basis).
We denote as $\varpi_r$ the corresponding projectors:
\begin{equation}\label{C0}
\varpi_r=\ket{X;r}\bra{X;r};\;\;\;\sum_r \varpi_r={\bf1}.
\end{equation}
Through a Fourier transform $F$ we get another orthonormal basis that we call momentum basis:
\begin{equation}
|{P};r\rangle=F|{X};r\rangle;\;\;\;\;
F=\frac{1}{\sqrt d}\sum _{r,s}\omega^{rs}\ket{X;r}\bra{X;s};\;\;\;\;
\omega=\exp \left (i\frac{2\pi }{d}\right );\;\;\;r,s\in {\mathbb Z}_d.
\end{equation}
The displacement operators $Z^\alpha, X^\beta$ in the phase space ${\mathbb Z}_d\times {\mathbb Z}_d$, are given by
\begin{eqnarray}\label{2A}
&&Z^\alpha= \sum \omega^{\alpha m}|X; m\rangle\bra{X;m}=\sum |P; m+\alpha \rangle \bra{P;m}\nonumber\\
&&X^\beta =\sum |X; m+\beta\rangle \bra{X; m}= \sum \omega^{-m\beta}|P; m \rangle \bra{P;m}.
\end{eqnarray}
where $\alpha$, $\beta \in {\mathbb Z}_d$. 
General displacement operators are the unitary operators
\begin{equation}\label{AA1}
D(\alpha, \beta)=Z^\alpha X^\beta \omega^{-2^{-1}\alpha \beta};\;\;\;\alpha, \beta \in {\mathbb Z}_d.
\end{equation}

Coherent states are defined as
\begin{equation}\label{bba}
\ket{\alpha, \beta}=D(\alpha, \beta)\ket{g},
\end{equation}
where $\ket{g}$ is a fiducial vector, different from position or momentum states so that we get a non-trivial set of $d^2$ states.
The corresponding projectors  ${\cal P}(\alpha, \beta)$ obey the resolution of the identity:
\begin{equation}\label{C1}
\frac{1}{d}\sum_{\alpha,\beta} {\cal P}(\alpha, \beta)={\bf 1};\;\;\;{\cal P}(\alpha, \beta)=\ket{\alpha, \beta}\bra{\alpha, \beta}.
\end{equation}
${\cal P}(\alpha, \beta)$ is related to ${\cal P}(\gamma, \delta)$ through a unitary transformation:
\begin{equation}
{\cal P}(\alpha, \beta)=D(\alpha-\gamma, \beta-\delta){\cal P}(\gamma, \delta)[D(\alpha -\gamma, \beta-\delta)]^{\dagger}.
\end{equation}

The projectors in Eqs(\ref{C0}),(\ref{C1}) can be used for the expansion of an arbitrary state $\ket{f}$ in terms of them:
\begin{eqnarray}\label{20}
&&\ket{f}=\sum_rf_r\ket{X;r};\;\;\;f_r=\bra{X;r}f\rangle\nonumber\\
&&\ket{f}=\frac{1}{d}\sum_{\alpha, \beta}f(\alpha, \beta)\ket{\alpha, \beta};\;\;\;f(\alpha, \beta)=\bra{\alpha, \beta}f\rangle.
\end{eqnarray}

An observable  is described with a $d\times d$ Hermitian matrix $\theta$.
Let $e_r(\theta)$ be the eigenvalues and $\Pi_r(\theta)$ the eigenprojectors of $\theta$.
Then
\begin{eqnarray}\label{99}
\theta=\sum _re_r(\theta)\Pi_r(\theta);\;\;\;\sum_r\Pi_r(\theta)={\bf 1};\;\;\;\Pi_r(\theta)\Pi_s(\theta)=\delta_{rs}\Pi_r(\theta);\;\;\;{\rm rank}[\Pi_r(\theta)]=1.
\end{eqnarray} 
Some eigenvalues might be zero, and the number of non-zero eigenvalues is equal to ${\rm rank}(\theta)$.

A measurement on a system described with the density matrix $R$, will give the result $e_r(\theta)$ with probability 
\begin{eqnarray}
p_r={\rm Tr}[R\Pi_r(\theta)];\;\;\;\sum_r p_r=1.
\end{eqnarray} 
The expectation value of $\theta$ is
\begin{eqnarray}
\langle\theta\rangle={\rm Tr}(R\theta)=\sum _re_r(\theta){\rm Tr}[R\Pi_r(\theta)].
\end{eqnarray} 
For simplicity the density matrix is not indicated in the notation, but it will be clear from the context.

\section{A partial order for  Hermitian matrices}\label{sec37}
We use the L\"{o}wner partial order \cite{L1,L2} among the Hermitian matrices (observables) and study some of its properties. 
The formalism can be viewed as a generalisation for partially ordered sets, of cumulative quantities which are used extensively with scalars in totally ordered sets.

\begin{definition}\label{100}
If $\theta, \phi \in M$, then  $\theta\succ \phi$ if $\theta-\phi$ is a positive semi-definite matrix (equivalently if $e_1(\theta-\phi)\ge0$).
If either $\theta\succ \phi$ or $\theta\prec \phi$, the matrices are comparable (and $\theta-\phi\in M_{P}$ or $\theta-\phi\in M_{N}$). 
If neither $\theta\succ \phi$ nor $\theta\prec \phi$, the matrices are non-comparable (and $\theta-\phi\in M_{PN}$).
\end{definition}

A set of $n$ Hermitian matrices such that
\begin{eqnarray}
\theta_1\prec\theta_2\prec...\prec\theta_n,
\end{eqnarray} 
is called a cumulative chain of matrices.

\begin{proposition}\label{proA1}
$\prec$ is a partial order in $M$.
\end{proposition}
\begin{proof}
It is easily seen that reflexivity, antisymmetry and transitivity hold:
\begin{eqnarray}
&&\theta \prec \theta,\nonumber\\
&&\theta \prec \phi\;{\rm and}\;\phi \prec \theta\;\Rightarrow\;\theta=\phi\nonumber\\
&&\theta \prec \phi\;{\rm and}\;\phi \prec \psi\;\Rightarrow\;\theta \prec \psi.
\end{eqnarray} 
\end{proof}

The following proposition shows that the L\"{o}wner partial order is `physically reasonable', in the sense  that
$\theta\succ \phi$ means that the expectation value for $\theta$ is greater than the expectation value for $\phi$, for all density matrices.

\begin{proposition}\label{proA2}
 $\theta\succ \phi$ if and only if for {\bf all} density matrices $R$,
\begin{eqnarray}
{\rm Tr}(R\theta)\ge {\rm Tr}(R\phi).
\end{eqnarray} 

\end{proposition}
\begin{proof}
Let $f_r, \Pi_r$ be the eigenvalues and eigenprojectors of $\theta-\phi$. Then 
\begin{eqnarray}
{\rm Tr}[R(\theta-\phi)]=\sum _rf_r{\rm Tr}(R\Pi_r).
\end{eqnarray} 
If $\theta-\phi$ is positive-semidefinite matrix, then all $f_r\ge0$ and the ${\rm Tr}[R(\theta-\phi)]\ge 0$.
Conversely, if ${\rm Tr}[R(\theta-\phi)]\ge 0$ for all $R$, we choose $R=\Pi_s$ and we prove $f_s\ge 0$ for all $s$.
Therefore $\theta-\phi$ is positive-semidefinite matrix.

\end{proof}

It is easily seen that 
\begin{itemize}
\item[(1)]
\begin{eqnarray}
\theta \prec \phi\;\Leftrightarrow \theta+\psi\prec \phi+\psi.
\end{eqnarray} 
\item[(2)]
If $\theta_1\prec\phi_1$ and $\theta_2\prec\phi_2$, then
 \begin{eqnarray}
\lambda_1\theta_1+ \lambda_2\theta_2\prec \lambda_1\phi_1+ \lambda_2\phi_2;\;\;\;\lambda_1,\lambda_2\ge 0.
\end{eqnarray} 

\item[(3)]
If $\theta \prec \phi$ and $0\le \lambda\le 1$ then
\begin{eqnarray}
\theta \prec (1-\lambda)\theta+\lambda\phi\prec \phi.
\end{eqnarray} 
Therefore as $\lambda$ takes values from $0$ to $1$, we get a chain of cumulative matrices.
\item[(4)]
If $U$ is a unitary operator, then
\begin{eqnarray}
\theta \prec \phi\;\Leftrightarrow\;U\theta U^\dagger\prec U\phi U^\dagger.
\end{eqnarray} 
But we see below (proposition \ref{L0}) that if $\theta \prec \phi$ there exist unitary transformations $U_1\ne U_2$ such that the $U_1\theta U_1^\dagger$ and $U_2\phi U_2^\dagger$ are non-comparable.

\item[(5)]
\begin{eqnarray}\label{23}
\theta \prec \phi\;\Rightarrow\;{\rm Tr}(\theta)\le {\rm Tr}(\phi).
\end{eqnarray} 

\end{itemize}

We next show that $\theta\succ \phi$ implies that $e_r (\theta)\ge e_r (\phi)$, but the converse is not true.
Therefore {\bf $\theta\succ \phi$ is stronger statement than $e_r (\theta)\ge e_r (\phi)$}.
Only for commuting matrices $e_r(\theta-\phi)=e_r (\theta)- e_r (\phi)$, and then $\theta\succ \phi$  is equivalent statement to $e_r (\theta)\ge e_r (\phi)$.

The following proposition (based on the Courant theory) is known in linear algebra (e.g., \cite{M} p.301), and we give it without proof.
\begin{proposition}\label{L10}
If $\theta\succ \phi$ , then :
\begin{itemize}
\item[(1)]
 The eigenvalues of $\theta, \phi$ obey the inequalities $e_r (\theta)\ge e_r (\phi)$.
\item[(2)]
If ${\rm rank}(\theta-\phi)=n$ then
$e_r(\theta)\le e_{r+n}(\phi)$ for $r=1,...,d-n$.
\end{itemize}

\end{proposition}
The following proposition shows that for any Hermitian matrix $\phi$ with eigenvalues $e_r(\phi)$ and any $f_r\ge e_r(\theta)$ for all $r$, we can always find another 
greater (according to the L\"{o}wner partial order) matrix with eigenvalues the $f_r$.
But not every matrix with these eigenvalues $f_r$, is comparable to $\phi$
(the converse of the first statement in the above proposition is not true).

\begin{proposition}\label{L0}
\mbox{}
\begin{itemize}
\item[(1)]
Let
\begin{eqnarray}
\phi=\sum_re_r(\phi)\Pi_r(\phi).
\end{eqnarray}
If $f_r\ge e_r(\phi)$ for all $r$, then
\begin{eqnarray}
\theta=\sum_rf_r\Pi_r(\phi)\succ \phi;\;\;\;[\theta, \phi]=0.
\end{eqnarray}
\item[(2)]
Let $\theta\succ \phi$ (in which case $e_r (\theta)\ge e_r (\phi)$). If ${\rm rank}(\theta-\phi)\le d-1$, 
there exists  unitary transformations $U_1\ne U_2$ such that the $\theta^\prime=U_1\theta U_1^\dagger$ and $\phi^\prime=U_2\phi U_2^\dagger$ 
are not comparable (although $e_r (\theta^\prime)\ge e_r (\phi^\prime)$).
\end{itemize}
\end{proposition}

\begin{proof}
\mbox{}
\begin{itemize}
\item[(1)]
The proof of this is straightforward.
\item[(2)]
The interval ${\cal I}(\theta)=[e_1(\theta),e_d(\theta)]$ is the set of all values of 
$\sum _r x_r^*(U_1\theta U_1^\dagger)_{rs}x_s$, for all unitary matrices $U_1$ and a fixed vector $x$ with $||x||=1$.
Also the interval ${\cal I}(-\phi)=[-e_d(\phi),-e_1(\phi)]$ is the set of all values of 
$\sum _r x_r^*(-U_2\phi U_2^\dagger)_{rs}x_s$, for all unitary matrices $U_2$ and a fixed vector $x$ with $||x||=1$.

It follows that for all unitary transformations $U_1,U_2$ we get all values in the interval
\begin{eqnarray}\label{10}
e_1(\theta)-e_d(\phi)\le\sum _{r,s} x_r^*(\theta^\prime_{rs}-\phi^\prime_{rs})x_s\le e_d(\theta)-e_1(\phi);\;\;\;||x||=1.
\end{eqnarray}
If ${\rm rank}(\theta-\phi)\le d-1$, the $e_1(\theta)-e_d(\phi)\le 0$ (proposition \ref{L10}).
Therefore there exists  unitary transformations $U_1, U_2$ for which $\sum  x_r^*(\theta^\prime_{rs}-\phi^\prime_{rs})x_s$ is negative, and 
consequently the $\theta^\prime \succ \phi^\prime$ does not hold (although $e_r (\theta^\prime)\ge e_r (\phi^\prime)$).
\end{itemize}
 \end{proof}

\subsection{Examples of non-comparable physical quantities}\label{sec100}

In this section we show that any two of the projectors $\varpi_r$ (in Eq.(\ref{C0})), and any two of the projectors ${\cal P}(\alpha, \beta)$ (in Eq.(\ref{C1})), are non-comparable.
They are used in expansions of an arbitrary state in terms of them (Eq.(\ref{20})).
Intuitively we expect that operators used in expansions are independent from each other (e.g., the projectors $\varpi_r$)
or partly independent from each other (e.g., the projectors ${\cal P}(\alpha, \beta)$) and this is related to the concept of non-comparable.
In addition to that we show that other physical quantities (like any two density matrices) are non-comparable.

The present section differentiates the various approaches that use these quantities, from our approach which uses cumulative (comparable) quantities.

\begin{proposition}\label{67}
Let $\theta$ be a Hermitian matrix, and $U$ a unitary matrix which does not commute with $\theta$.
Then $\theta$ and $U\theta U^\dagger$ are non-comparable.
\end{proposition}
\begin{proof}
Since ${\rm Tr}(\theta)-{\rm Tr}(U\theta U^\dagger)=0$ it follows that $\theta-U\theta U^\dagger$ has both positive and negative eigenvalues, and therefore
the $\theta$ and $U\theta U^\dagger$ are non-comparable.
\end{proof}

From proposition \ref{67} follows immediately the next corollary:
\begin{cor}\label{cor1}
\mbox{}
\begin{itemize}
\item[(1)]
Any pair of projectors $\varpi_r, \varpi_s$ (defined in Eq.(\ref{C0})) are non-comparable.
\item[(2)]
Any pair of coherent projectors ${\cal P}(\alpha_1,\beta_1)$, ${\cal P}(\alpha_2,\beta_2)$ (defined in Eq.(\ref{C1})), are non-comparable.
\item[(3)]
 Any pair of (normalised) density matrices $R_1, R_2$, are non-comparable.
 \end{itemize}
\end{cor} 
 \begin{proof}
 The first two statements follow immediately from proposition \ref{67}.
 
 For the third statement we note that if
$R_1\prec R_2$, that would imply $e_r(R_1)\le e_r(R_2)$ for their eigenvalues. Since
${\rm Tr}(R_1)={\rm Tr}(R_2)=1$ we would have $e_r(R_1)= e_r(R_2)$, i.e., $R_1=UR_2U^\dagger$ for some unitary operator $U$. But according to proposition \ref{67} this contradicts the fact that  $R_1,R_2$ are comparable.
\end{proof}
The third statement in this corollary is related to proposition \ref{pro38} below (in the topological language).

\section{Upper and lower sets: cumulative matrices in the partially ordered set $M$}\label{sec101}

Cumulativeness for scalars in totally ordered sets is a practical concept, which has been applied extensively in probability theory, Mathematical Economics, Finance, etc. 
It leads to a chain with increasing  values of some scalar quantity.

In partially ordered sets we have  non-comparable quantities, and cumulativeness needs to be generalised.
In this section we do this by using upper and lower sets within  the partially ordered set $M$ of Hermitian matrices.
Later we get a deeper understanding of their properties using the language of topology, and apply these ideas to Choquet integrals.
\begin{definition}
\mbox{}
\begin{itemize}
\item
${\cal U}\subseteq M$ is an upper set if $\theta\in {\cal U}$ and $\phi\succ \theta$, implies $\phi \in {\cal U}$. 
Physically this means that if $\theta\in {\cal U}$, then the observable $\phi \in {\cal U}$ if and only if ${\rm Tr}(R\theta)\le {\rm Tr}(R\phi)$ for all density matrices $R$.
\item
${\cal L}\subseteq M$ is a lower set if $\theta\in {\cal L}$ and $\phi\prec \theta$, implies $\phi \in {\cal L}$.
Physically this means that if $\theta\in {\cal L}$ , then the observable $\phi \in {\cal L}$ if and only if ${\rm Tr}(R\theta)\ge {\rm Tr}(R\phi)$ for all density matrices $R$.
\end{itemize}
\end{definition}
\begin{remark}
\mbox{}
\begin{itemize}
\item[(1)]
The complement of an upper set is a lower set. The union and intersection of two upper sets, are upper sets. The same is true for lower sets.
This is important later, because it implies that we can use the language of topology for these sets.
\item[(2)]
The $\emptyset, M$ are both upper and lower sets.
$M_P$ and $M_{PN}\cup M_P$ are upper sets.
$M_N$ and $M_{PN}\cup M_N$ are lower sets.
$M_{PN}$ is neither upper, nor lower set.
\item[(3)]
The upper sets contain many infinite cumulative chains of matrices, but they also contain non-comparable matrices (e.g, example \ref{ex23} below).
The same is true for lower sets. In this sense upper and lower sets in partially ordered structures, are generalisations of the (upwards and downwards) cumulative chains in totally ordered structures.
\item[(4)]
Let $A\subseteq M$ and ${\mathfrak U}(A)$ be a set such that for all $\theta\in A$, if $\phi\succ \theta$  then $\phi \in {\mathfrak U}(A)$. 
In general ${\mathfrak U}(A)$ is {\bf not} an upper set (because if $\phi\succ \theta$ and $\theta\in {\mathfrak U}(A)\setminus A$, there is no requirement that $\phi \in{\mathfrak U}(A)$).
\end{itemize}
\end{remark}
\begin{definition}
An upper set ${\cal U}$ is directed upwards,  if for every pair $\theta, \phi\in{\cal U}$ there exists $\psi\in{\cal U}$  such that $\theta\prec \psi$ and $\phi\prec\psi$.
In analogous way we define lower sets which are directed downwards.
\end{definition}  
\begin{lemma}\label{L12}
The upper sets defined above are upwards directed sets. The lower sets defined above are downwards directed sets. $M$ is a directed set (both upwards and downwards). 
\end{lemma}
\begin{proof}
Let ${\cal U}$ be an upper set.
We express the Hermitian matrices $\theta, \phi\in {\cal U}$ in terms of their eigenvalues $e_r(\theta), e_r(\phi)$ and eigenprojectors $\Pi_r(\theta), \Pi_r(\phi)$, correspondingly:
\begin{eqnarray}
\theta=\sum e_r(\theta)\Pi_r(\theta);\;\;\;\phi=\sum e_r(\phi)\Pi_r(\phi).
\end{eqnarray} 
Let
\begin{eqnarray}
\widehat \theta=\sum |e_r(\theta)|\Pi_r(\theta)\succ \theta;\;\;\;\widehat \phi=\sum |e_r(\phi)|\Pi_r(\phi)\succ \phi.
\end{eqnarray} 
Clearly $\widehat \theta, \widehat \phi\in{\cal U}$ and they are positive semi-definite.
Then there exists $\psi$  such that $\psi\succ\widehat \theta$ and $\psi\succ \widehat \phi$.
Examples of this are all
$\psi=\lambda \widehat \theta+\mu \widehat \phi$ with $\lambda, \mu\ge 1$.
It follows that there exists $\psi$ such that $\theta\prec \psi$ and $\phi\prec\psi$.

In analogous way we prove that lower sets are directed set downwards. 
$M$ is both lower and upper set, and therefore it is is a directed set (both upwards and downwards).
\end{proof}

\begin{lemma}\label{L1}
Any two upper sets (or any two lower sets) cannot be disjoint.
\end{lemma}
\begin{proof}
Let ${\cal U}_1$ and ${\cal U}_2$ be two upper sets.
Also let $\theta_1\in {\cal U}_1$ and $\theta_2 \in {\cal U}_2$.
Since $M$ is a directed set there exists a matrix $\psi$ such that $\psi\succ\theta _1$ and $\psi\succ\theta _2$, and therefore $\psi$ belongs to both ${\cal U}_1$ and ${\cal U}_2$.

In a similar way we prove the statement for two lower sets.
\end{proof}

\subsection{Examples of upper and lower sets in $M$}\label{sec29}

The examples in this section, are needed in section \ref{topo} where we use the topological language.
\begin{example}
The set 
\begin{eqnarray}\label{4}
{\cal O}(\theta)=\{\theta+\rho\;|\;\rho \in M_P\},
\end{eqnarray} 
is an upper set. A special case of this is the ${\cal O}(0)=M_P$. Then:
\begin{itemize}
\item[(1)]
If $\theta$ is a density matrix $R$, the ${\cal O}(R)$ contains no other density matrix.
This is because ${\rm Tr}(R)=1$ and ${\rm Tr}(R+\rho)>1$ for any $\rho\in M_P$. Therefore
\begin{eqnarray}\label{46A}
{\cal O}(R)\cap {\mathfrak D}=\{R\};\;\;\;R\in{\mathfrak D}.
\end{eqnarray} 
This result is needed later.

\item[(2)]
For $\theta\ne 0$, the ${\cal O}(\theta)$ does not contain all matrices in $M_P$. 
We show that if $e_r(\theta)\le f_r$, there are matrices with eigenvalues the $f_r$ which belong to  ${\cal O}(\theta)$,
but there are other matrices with the same eigenvalues $f_r$ which do not belong to  ${\cal O}(\theta)$.

We express $\theta$ in terms of its eigenvectors $e_r(\theta)$ and eigenprojectors $\Pi_r(\theta)$, and let $f_r\ge e_r(\theta)$:
\begin{eqnarray}
\theta=\sum e_r(\theta)\Pi_r(\theta);\;\;\;f_r\ge e_r(\theta).
\end{eqnarray} 
Then there are matrices $\phi$ which have the $f_r$ as eigenvalues such that $\phi\succ \theta$. An example of this is 
\begin{eqnarray}
\phi=\sum f_r\Pi_r(\theta);\;\;\;f_r\ge e_r(\theta).
\end{eqnarray} 
Equally there are matrices $\psi=U\phi U^\dagger$ where $U$ is some unitary transformation, which have the $f_r$ as eigenvalues, and are not comparable to $\theta$ (proposition \ref{L0}).
\item[(3)]
Its complement $M\setminus {\cal O}(\theta)$ is a lower set.
\end{itemize}

In a similar way the set 
\begin{eqnarray}\label{5}
{\cal C}(\phi)=\{\phi-\rho\;|\;\rho \in M_P\},
\end{eqnarray}
is a lower set. A special case of this is the ${\cal C}(0)=M_N$.
Its complement $M\setminus {\cal C}(\phi)$ is an upper set.
For later use, we note that for a density matrix $R$, the analogue of Eq.(\ref{46A}) is
\begin{eqnarray}\label{47A}
{\cal C}(R)\cap {\mathfrak D}=\{R\};\;\;\;R\in{\mathfrak D}.
\end{eqnarray} 
\end{example}
\begin{example}\label{ex23}
We consider the set
\begin{eqnarray}
{\cal O}(\theta_1)\cup {\cal O}(\theta_2)=\{\theta_1+\rho_1\;{\rm or}\;\theta_2+\rho_2|\rho_1,\rho_2\in M_P\},
\end{eqnarray}
which as union of upper sets , is an upper set.
If $\theta_1\prec\theta_2$, then ${\cal O}(\theta_1)\cup {\cal O}(\theta_2)={\cal O}(\theta_1)$.
If $\theta_1, \theta_2$ are non-comparable, then ${\cal O}(\theta_1)\cup {\cal O}(\theta_2)$ is different from the
${\cal O}(\theta_1)$ and ${\cal O}(\theta_2)$.
For later use we note that if $\theta_1$ is a density matrix $R_1$ and $\theta_2$ is a density matrix $R_2$, then similar argument with the previous example shows that
\begin{eqnarray}\label{46B}
[{\cal O}(R_1)\cup {\cal O}(R_2)]\cap {\mathfrak D}=\{R_1,R_2\};\;\;\;R_1,R_2\in{\mathfrak D}.
\end{eqnarray} 

We also consider the set
${\cal C}(\theta_1)\cup {\cal C}(\theta_2)$ which as union of lower sets , is a lower set.
If $\theta_1\prec\theta_2$, then ${\cal C}(\theta_1)\cup {\cal C}(\theta_2)={\cal C}(\theta_2)$.
If $\theta_1$ is a density matrix $R_1$ and $\theta_2$ is a density matrix $R_2$, then similar argument with the previous example shows that
\begin{eqnarray}\label{47B}
[{\cal C}(R_1)\cup {\cal C}(R_2)]\cap {\mathfrak D}=\{R_1,R_2\};\;\;\;R_1,R_2\in{\mathfrak D}.
\end{eqnarray} 
\end{example}
\begin{example}\label{ex24}

The previous example can be generalised to a union of many upper sets ${\cal O}[\theta(\lambda)]$:
\begin{eqnarray}
\bigcup_{\lambda\in \Lambda}{\cal O}[\theta(\lambda)]=\{\theta(\lambda)+\rho(\lambda)\;{\rm for\; all}\;\lambda\in\Lambda|\rho(\lambda)\in M_P\}.
\end{eqnarray} 
$\lambda$ is a continuous or discrete parameter that takes values in a set $\Lambda$.
An example of this is to take the unitary time evolution relation $\theta(\lambda)=\exp(i\lambda {\mathfrak H})\theta_0\exp(-i\lambda {\mathfrak H})$, where $\lambda$ is time and ${\mathfrak H}$ is a Hamiltonian.
In this case $\Lambda$ is the set of real numbers.

If all $\theta(\lambda)$ are density matrices $R(\lambda)$, then
\begin{eqnarray}\label{46C}
\left [\bigcup_{\lambda\in \Lambda}{\cal O}[R(\lambda)]\right ]\cap {\mathfrak D}=\bigcup_{\lambda\in \Lambda}[{\cal O}[R(\lambda)]\cap {\mathfrak D}]
=\{R(\lambda)|\lambda\in\Lambda\};\;\;\;R(\lambda)\in{\mathfrak D}.
\end{eqnarray} 
Also for a union of many lower sets ${\cal C}[R(\lambda)]$ we find
\begin{eqnarray}\label{47C}
\left [\bigcup_{\lambda\in \Lambda}{\cal C}[R(\lambda)]\right ]\cap {\mathfrak D}=\bigcup_{\lambda\in \Lambda}[{\cal C}[R(\lambda)]\cap {\mathfrak D}]=\{R(\lambda)|\lambda\in\Lambda\};\;\;\;R(\lambda)\in{\mathfrak D}.
\end{eqnarray} 
It is important later, that the sets in Eqs(\ref{46C}), (\ref{47C}) are the same (and also in Eqs(\ref{46B}), (\ref{47B}), and also in Eqs(\ref{46A}), (\ref{47A})).

\end{example}

\section{The topological space of Hermitian matrices and observables}\label{topo}

In the previous section we linked (upwards and downwards) cumulativeness in partially ordered structures, with upper and lower sets.
In this section we get a deeper understanding of their properties using the language of topology, which we link with the physical language.

We first recall briefly some of the topological concepts, which are needed below\cite{TT1,TT2}.
\begin{itemize}
\item[(1)]
A topological space $(X, {\cal T}_X)$ is a set $X$ of `points' and a collection ${\cal T}_X$ of its subsets (called open sets). We require that
 every union of elements of ${\cal T}_X$ belongs in ${\cal T}_X$, and every intersection of a finite number of elements of ${\cal T}_X$ belongs in ${\cal T}_X$.
The $\emptyset, X$ are also elements of ${\cal T}_X$. The complements of open sets are called closed sets.
The $\emptyset, X$ are both open and closed sets. 
\item[(2)]
An open base ${\cal B}\subseteq {\cal T}_X$ is a set of open sets such that every open set is a union of the open sets in the base.
\item[(3)]
A neighbourhood of a point $x$, is a subset of $X$ that contains an open set that contains the point $x$.
A  fundamental system of neighbourhoods of a point $x$, is a set ${\cal G}$ of neighbourhoods such that for each neighbourhood ${\cal N}_1$ of the point $x$,
there is a neighbourhood ${\cal N}_2\in {\cal G}$, such that ${\cal N}_2\subseteq{\cal N}_1$. In some examples ${\cal G}$ has only one element, and this is the case below.

\item[(4)]
Let $Y\subseteq X$. The topology induced on Y by the topology on X, has as open (closed) sets the intersection of the open (closed) sets in $X$ with $Y$.
We then say that $Y$ is a subspace of $X$.
We note that an open (closed) set in the subspace $Y$, might {\bf not} be open (closed) in $X$.
Also we need to distinguish carefully between the properties of points of the subspace $Y$, and the properties of the same points as points of $X$ (this is important below).
\end{itemize}

In our context we take $X$ to be the set $M$ of Hermitian matrices, and ${\cal T}_M$ to be the set of all the upper sets ${\cal U}$.
It is easily seen  that the union and finite intersection of upper sets, is an upper set.
Therefore $(M, {\cal T}_M)$ is a  topological space where:
\begin{itemize}
\item[(1)]
The points are the Hermitian matrices (observables).
\item[(2)]
The open sets are the upper sets ${\cal U}$. They satisfy the requirement that every union and every finite intersection of upper sets, is an upper set 
Physically if an observable belongs to ${\cal U}$, then any other greater (according to the L\"{o}wner order) observable, also belongs to ${\cal U}$.

The sets $M_P, M_P\cup M_{PN}, M,\emptyset$, ${\cal O}(\theta)$, are examples of open sets.
According to lemma \ref{L1} there are no disjoint open sets in $M$. A neighbourhood of $\theta$ is a set that contains any open set that contains $\theta$ (e.g, it contains the ${\cal O}(\theta)$).
\item[(3)]
The closed sets are the lower sets ${\cal L}$ (complements of the open sets).
Physically if an observable belongs to ${\cal L}$, then any other smaller (according to the L\"{o}wner order) observable also belongs to ${\cal L}$.

The sets $M_N, M_N\cup M_{PN}, M,\emptyset$, ${\cal C}(\theta)$, are examples of closed sets.
\item[(4)]
The set of all ${\cal O}(\theta)$ (Eq.(\ref{4})) for all $\theta\in M$, is an open basis in $M$. 
All open sets are unions of these sets, and some examples of this have been given in examples \ref{ex23}, \ref{ex24}.

\end{itemize}

\subsection{Separation axioms}\label{secA5}

From the separation axioms point of view, the set of 
$T_2$-spaces (Hausdorff spaces) is a subset of the set of $T_1$-spaces, i.e.,  Hausdorff spaces have stronger properties than $T_1$-spaces. 
Also the set of $T_1$-spaces (Frechet spaces) is a subset of the set of $T_0$-spaces (Kolmogorov spaces).
In Analysis we usually have Hausdorff spaces, while $T_1$ and $T_0$ spaces are usually used in connection with logic.

In a very different from the present paper physical context, we have used $T_0$ topologies to describe the logical relationship between a quantum system and its subsystems\cite{T1,T2}.
Here we show that the topological space $(M, {\cal T}_M)$ of Hermitian matrices (observables) is a  ${T_0}$-space, and it is not $T_1$-space.
We also discuss the physical significance of this.

\begin{proposition}
$(M, {\cal T}_M)$ is a  ${T_0}$-space, and it is {\bf not} $T_1$-space (and therefore it is not Hausdorff space).
\end{proposition}
\begin{proof}
\mbox{}
\begin{itemize}
\item[(1)]
A topological space is $T_0$-space if for all distinct points $\theta, \phi$ there exists an open set ${\cal U}$, such that either $\theta \in {\cal U}$ and $\phi \notin {\cal U}$ or 
$\phi \in {\cal U}$ and $\theta \notin {\cal U}$. We have the following cases:
\begin{itemize}
\item
If $\theta \prec\phi$ then we take the open set ${\cal O}(\phi)$. Clearly $\theta\notin {\cal O}(\phi)$.
\item
The case $\phi \prec \theta$ is analogous to the previous one.
\item
If $\theta, \phi$ are not comparable (i.e., $\theta-\phi\in M_{PN}$), then ${\cal U}$ can be taken to be the open set that contains the Hermitian matrices  $\phi+\rho$ for all $\rho\in M_P$. Clearly $\theta$ does not belong to ${\cal U}$.
\end{itemize}
Therefore $(M, {\cal T}_M)$ is a $T_0$-space.

\item[(2)]
A topological space is $T_1$-space if for all distinct points $\theta, \phi$ there exist two open sets ${\cal U}_1$ and ${\cal U}_2$, such that $\theta \in {\cal U}_1$, $\phi \in {\cal U}_2$ and
 $\theta \notin {\cal U}_2$, $\phi \notin {\cal U}_1$. If $\theta \prec \phi$ this is not possible, because any open set that contains $\theta$ also contains $\phi$.
Therefore $(M, {\cal T}_M)$ is not $T_1$-space.

\item[(3)]
$M$ is not a $T_1$-space, and therefore it is not a Hausdorff space.
A topological space is Hausdorff space ($T_2$-space) if for all distinct points $\theta, \phi$ there exist two {\bf disjoint} open sets ${\cal U}_1$ and ${\cal U}_2$, such that 
$\theta \in {\cal U}_1$, $\phi \in {\cal U}_2$ and $\theta \notin {\cal U}_2$, $\phi \notin {\cal U}_1$.
We have seen in lemma \ref{L1} that in $M$ there are no disjoint open (upper) sets.
\end{itemize}
\end{proof}
{\bf Physical meaning:}
The topological concept of a $T_1$-space is incompatible with the physical concept of cumulativeness.
In a partially ordered set, there exist $\theta,\phi$ which are in the same chain and therefore in the same open set.
Therefore we cannot satisfy the requirement that  for all $\theta, \phi$ there exist two open sets such that one of them contains $\theta$ but not $\phi$, and the other contains $\phi$ but not $\theta$.
So from a physical point of view it not surprising that the topological space $M$ is not $T_1$-space.

\subsection{Connectedness}

Here we discuss the connectedness of the topological space $(M, {\cal T}_M)$.
We do not discuss compactness, because most authors define compactness only for Hausdorff spaces (e.g., Bourbaki\cite{TT1}).
$M$ is not a Hausdorff space and for this reason we do not discuss compactness.

\begin{proposition}
The topological space $(M, {\cal T}_M)$ is both connected and  locally connected.
\end{proposition}

\begin{proof}
For connectedness we need to prove  that $M$ cannot be written as the union of two disjoint open sets.
According to lemma \ref{L1} there are no disjoint open (upper) sets, and therefore 
this proves that $M$ is connected.

For every point (matrix) $\theta$, the ${\cal O}(\theta)$ (Eq.(\ref{4})) is a subset of every neighbourhood of $\theta$, and therefore the $\{{\cal O}(\theta)\}$ (which contains only one open set) is a 
fundamental system of neighbourhoods. The ${\cal O}(\theta)$ is connected, and therefore the topological space $(M, {\cal T}_M)$ is locally connected.
\end{proof}

\subsection{Topology induced on the subset ${\mathfrak D}$ of density matrices}\label{secA6}

The topology in $M$ induces a topology in its subset ${\mathfrak D}$ of density matrices.
The open (closed) sets in ${\mathfrak D}$, are the intersections of the the open (closed) sets in $M$ with ${\mathfrak D}$.
Examples of open sets in ${\mathfrak D}$ (they are {\bf not} open sets in $M$) are in Eqs(\ref{46A}), (\ref{46B}), (\ref{46C}):
\begin{eqnarray}
\{R\};\;\;\;\{R_1,R_2\};\;\;\;\{R(\lambda)|\lambda\in\Lambda\};\;\;\;R,R_1,R_2, R(\lambda)\in{\mathfrak D}.
\end{eqnarray}
As we discussed in example \ref{ex24}, $\lambda$ is a discrete or continuous variable (e.g., time in unitary time evolution of a density matrix).
We note that the same sets are also examples of closed sets in ${\mathfrak D}$ as can be seen in Eqs(\ref{47A}), (\ref{47B}), (\ref{47C}).

It is seen that the open sets in ${\mathfrak D}$ are also closed sets.
This is a characteristic of a totally disconnected topological space, and we elaborate on this in the proposition below.

\begin{proposition}\label{pro38}
The topological space of density matrices ${\mathfrak D}$, is Hausdorff and totally disconnected.
\end{proposition}
\begin{proof}
If $R_1, R_2$ are any distinct density matrices, then $\{R_1\}$ and $\{R_2\}$ are open sets in ${\mathfrak D}$ and 
\begin{eqnarray}
R_1\in \{R_1\};\;\;\;R_2\in \{R_2\};\;\;\;R_1\notin \{R_2\};\;\;\;R_2\notin \{R_1\}.
\end{eqnarray} 
Therefore the topological space ${\mathfrak D}$ is Hausdorff (in contrast to $M$ which is $T_0$-space).

We next show that any open set in ${\mathfrak D}$ with more than one element is not connected, i.e., it can be written as the union of disjoint open sets.
Indeed the open set $\{R_1,R_2\}$ is the union of the open sets $\{R_1\}$ and $\{R_2\}$. More generally a general open set in ${\mathfrak D}$ can be written as the union of open sets which have only one element.
We rewrite Eq.(\ref{46C}) as:
\begin{eqnarray}
\{R(\lambda)|\lambda\in\Lambda\}=\bigcup _{\lambda\in \Lambda}\{R(\lambda)\};\;\;\;R(\lambda)\in{\mathfrak D}.
\end{eqnarray} 
It is seen that the connected component of every point $R(\lambda)$ in ${\mathfrak D}$, is $\{R(\lambda)\}$ and has only one element.
This completes the proof.
\end{proof}

\begin{remark}\label{rem15}
The statement that ${\mathfrak D}$ is totally disconnected (in the topological language), is related to the statement that any two density matrices are non-comparable (in the partial order language in \ref{cor1}).
Indeed the statement that there exist open sets of density matrices that have only one element (the ${\cal O}(R)\cap {\mathfrak D}$), is linked to the statement that any two density matrices are non-comparable. 
Below (remark\ref{rem18}) we link these two statement to capacities and quantum Choquet integrals. 
\end{remark}

\section{Quantum Choquet integrals}  

\subsection{Capacities and Choquet integrals in a classical context}\label{sec13}

Let $\Omega$ be the set of all possible outcomes of an experiment (sample space) which we take to be finite:
\begin{eqnarray}\label{27}
\Omega=\{1,..,n\}.
\end{eqnarray}

Also let $2^\Omega$ be the powerset of $\Omega$ (the set of all $2^n$ subsets of $\Omega$).
Kolmogorov probability is a map $\mu$ from subsets of $\Omega$ to $[0,1]$, such that $\mu (\emptyset )=0$, $\mu (\Omega)=1$, and also
\begin{eqnarray}\label{1Z}
\mu(A\cup B)-\mu(A)-\mu(B)+\mu(A\cap B)=0;\;\;\;A,B \subseteq \Omega.
\end{eqnarray}
In the case $A\cap B=\emptyset$ this reduces to the additivity relation
\begin{eqnarray}\label{B}
A\cap B=\emptyset\;\;\rightarrow\;\;\mu (A\cup B)= \mu (A)+\mu (B).
\end{eqnarray}
Capacity or nonadditive probability, replaces Eq.(\ref{1Z}) with  the weaker relation
\begin{eqnarray}\label{A2}
A \subseteq B\;\rightarrow\;0\le \mu (A)\le \mu (B)\le 1
\end{eqnarray}
In this case even if $A\cap B=\emptyset$, in general we get 
\begin{eqnarray}\label{2AB}
\mu (A\cup B)\ne \mu (A)+\mu (B).
\end{eqnarray}
This expresses the fact that the `whole is not equal the sum of its parts'.

Since the additivity relation of Eq.(\ref{B}) does not hold for capacities, the concept of integration needs 
appropriate revision and this leads to Choquet integrals which are based on cumulative functions. For Kolmogorov probabilities, 
the derivatives of cumulative functions are  probability distributions. In the discrete case,
discrete derivatives  of cumulative functions are equal to probabilities (the proof of this is based on the additivity relation in Eq.(\ref{B})).
This is not true for capacities, and in this case there is merit in using cumulative functions. Choquet integrals are based on cumulative functions.

Let $f(r)$ (with $r=1,...,n$) be a function from $\Omega$ to ${\mathbb R}^+$. We introduce a `ranking permutation' $\sigma$ of the elements of $\Omega$ such that
\begin{eqnarray}\label{order}
0\le  f[\sigma (1)]\le f[\sigma (2)]\le...\le f[\sigma (n)].
 \end{eqnarray}
 
 Different functions are in general associated with different ranking permutations.
\begin{definition}
Functions with the same ranking permutation are called comonotonic. 
\end{definition}

The Choquet integral of $f$ with respect to the capacity $\mu$ is given by
\begin{eqnarray}\label{480}
{\cal C}(f)=\Delta f[\sigma (1)]\mu[\{\sigma (1),...,\sigma (n)\}]+\Delta f[\sigma (2)]\mu[\{\sigma(2)...,\sigma (n)\}]]+...+
\Delta f [\sigma (n)]\mu[\sigma (n)]
\end{eqnarray}
where $\Delta f [\sigma (r)]$ are the {\bf function increments}
\begin{eqnarray}
&&\Delta f [\sigma (r)]=f [\sigma (r)]-f [\sigma (r-1)]\ge 0;\;\;\;r=2,...,n\nonumber\\
&&\Delta f[\sigma (1)]=f[\sigma (1)]\ge 0.
\end{eqnarray}
It can also be written as
\begin{eqnarray}
{\cal C}(f)=f[\sigma (1)]\Delta \mu[\{\sigma (1),...,\sigma(\nu)\}]+f[\sigma (2)]\Delta\mu[\{\sigma(2),...,\sigma(\nu)\}]]+...+f [\sigma (n)]\Delta\mu[\sigma (n)]
\end{eqnarray}
where $\Delta \mu[\sigma (r)]$ are the {\bf capacity increments}
\begin{eqnarray}
&&\Delta \mu [\{\sigma (r),...,\sigma(\nu)\}]=\mu[\{\sigma (r),...,\sigma(\nu)\}]-\mu [\{\sigma (r-1),...,\sigma(n)\}]\ge 0;\;\;\;r=2,...,n\nonumber\\
&&\Delta \mu[\{\sigma (n)\}]=\mu [\{\sigma (n)\}]\ge 0.
\end{eqnarray}
We note here that in the special case that the additivity relation holds (Eq.(\ref{B})), then
\begin{eqnarray}
&&\Delta \mu [\{\sigma (r),...,\sigma(\nu)\}]=\mu[\{\sigma (r)\}].
\end{eqnarray}
In this special case ${\cal C}(f)$ is simply a weighted average of the $f[\sigma (r)]$ with weight $\mu[\{\sigma (r)\}]$.

It is easily seen that for comonotonic functions 
\begin{eqnarray}
{\cal C}(af_1+bf_2)=a{\cal C}(f_1)+b{\cal C}(f_2);\;\;\;a,b\ge 0,
\end{eqnarray}
but in general this is {\bf not true}.

Examples of the use of Choquet integrals in real life decision problems in a classical context, have been discussed in \cite{D1,D2,D3,D4,D5,D6,D7}.

\subsection{Capacities as expectation values of positive semi-definite matrices in an upper set}\label{sec14}

We consider the set $\Omega$ in Eq.(\ref{27}) and we map each of the $2^n$ subsets $A$ to a positive semi-definite $d\times d$ matrix (observable)  $\theta(A)\in M_P$ 
such that $\theta(\emptyset)={\bf 0}$, $\theta(\Omega)={\bf 1}$ and
\begin{eqnarray}\label{A}
A \subseteq B\;\rightarrow\;{\bf 0}\prec \theta(A)\prec\theta (B)\prec{\bf 1}.
\end{eqnarray}
In the topological language the ${\bf 0}, \theta(A), \theta (B), {\bf 1}$ belong to the open set ${\cal O}(0)=M_P$.
It follows that for {\bf all} density matrices $R$, the expectation values of these observables are capacities. Indeed 
\begin{eqnarray}\label{Q3}
A \subseteq B\;\rightarrow\;0\le\langle \theta(A)\rangle\le\langle \theta (B)]\rangle \le 1.
\end{eqnarray}

The upper sets introduced earlier, provide a systematic way to find matrices that obey Eq.(\ref{A}).
We choose arbitrarily the $\theta(\{1\}),...,\theta(\{n\})$ (where $0\prec \theta(\{r\})\prec{\bf 1}$ ). Then we consider the upper set (open set in the topological language)
\begin{eqnarray}
{\cal U}={\cal O}[\theta(\{1\})]\cup ...\cup{\cal O}[\theta(\{n\})].
\end{eqnarray}
Since ${\cal U}$ is an upwards directed set (lemma \ref{L12}), we can find many operators $\theta(r,s)$ such that
\begin{eqnarray}
\theta(\{r,s\})\succ \theta(\{r\});\;\;\;\theta(\{r,s\})\succ \theta(\{s\});\;\;\;r,s\in\Omega.
\end{eqnarray}
Examples are all operators $\theta(\{r,s\})=\mu_1\theta(\{r\})+\mu _2\theta(\{s\})$ with $\mu_1,\mu_2\ge 1$. 
This argument generalises to
\begin{eqnarray}
\theta(\{r,s,t\})\succ \theta(\{r,s\});\;\;\;\theta(\{r,s,t\})\succ \theta(\{r,t\});\;\;\;\theta(\{r,s,t\})\succ \theta(\{s,t\}), 
\end{eqnarray}
etc. 

\begin{lemma}
\begin{itemize}
\item[(1)]
For any $A\subseteq \Omega$
\begin{eqnarray}\label{56}
0\le {\rm Tr}[\theta (A)] \le d.
\end{eqnarray}
\item[(2)]
For any $A,B\subseteq \Omega$
\begin{eqnarray}\label{Q2}
0\le {\rm Tr}[\theta (A)\theta(B)] \le \min\{{\rm Tr}[\theta (A)], {\rm Tr}[\theta (B)] \}\le d.
\end{eqnarray}
\item[(3)]
\begin{eqnarray}\label{Q3A}
A \subseteq B\;\rightarrow\;0\le{\rm Tr}[\theta(A)]\le{\rm Tr}[ \theta (B)]\le d.
\end{eqnarray}

\end{itemize}

\end{lemma}
\begin{proof}
\begin{itemize}
\item[(1)]

We use Eq.(\ref{23}) together with the fact that ${\bf 0}\prec \theta(A)\prec {\bf 1}$, and we get Eq.(\ref{56}).

\item[(2)]
In $0\le \langle \theta (A)\rangle \le 1$ (Eq.(\ref{Q3})) we use as density matrix the $\frac{\theta(B)}{{\rm Tr}[\theta(B)]}$.
\begin{eqnarray}
0\le {\rm Tr}[\theta (A)\theta(B)] \le {\rm Tr}[\theta (B)] \le d.
\end{eqnarray}
From this follows Eq.(\ref{Q2}).

\item[(3)]
In Eq.(\ref{Q3}) we use as density matrix the $\frac{1}{d}{\bf 1}$ and we get Eq.(\ref{Q3A}).
\end{itemize}
\end{proof}

In general
\begin{eqnarray}\label{1}
\theta(A\cup B)-\theta(A)-\theta(B)+\theta(A\cap B)\ne0;\;\;\;A,B \subseteq \Omega.
\end{eqnarray}
From this follows that in the case $A\cap B=\emptyset$, in general
\begin{eqnarray}\label{66}
\theta (A\cup B)\ne \theta (A)+\theta (B).
\end{eqnarray}

\begin{remark}\label{rem18}
In Eq.(\ref{A}) the $ \theta(A), \theta (B)$ cannot both be density matrices.
In order to show the link between the section on partial order, the section on topology, and the present section, we explain this briefly in three different ways:
\begin{itemize}
\item
Any two density matrices are non-comparable (corollary \ref{cor1}).
\item
The set of density matrices is a totally disconnected topological space (see remark \ref{rem15}).
\item
We can prove directly that if $ \theta(A), \theta (B)$ are both density matrices, they cannot satisfy Eq.(\ref{Q3}) with all density matrices.
We consider two density matrices
\begin{eqnarray}
R_1=\sum_re_r\pi_r;\;\;\;R_2=\frac{1}{d-1}\sum_r(1-e_r)\pi_r=\frac{1}{d-1}({\bf 1}-R_1).
\end{eqnarray}
Here $\pi_r$ are their eigenprojectors and $e_r, \frac{1-e_r}{d-1}$ their eigenvalues.
Then we easily prove for the density matrices $ \theta(A), \theta (B), R_1, R_2$, that if ${\rm Tr}[R_1\theta(A)]< {\rm Tr}[R_1\theta(B)]$, then
${\rm Tr}[R_2\theta(A)]> {\rm Tr}[R_2\theta(B)]$.
Therefore the $ \theta(A), \theta (B)$ cannot  both be density matrices.
\end{itemize}

\end{remark}

\subsection{M\"obius transforms of positive semi-definite matrices}\label{sec15}
The M\"obius transform is used widely in Combinatorics,  after the work by Rota\cite{R}. It is a generalization of
the inclusion-exclusion principle in set theory which removes the overlaps (double-counting) between intersecting sets. Rota generalised this to partially ordered structures.
In the present context, the M\"obius transform of the positive semi-definite operators $\theta(A)$ is \cite{VV}
\begin{eqnarray}\label{M}
{\cal M}(A)=\sum  _{B\subseteq A}(-1)^{|A|-|B|} \theta(B).
\end{eqnarray}
Here $|A|$, $|B|$ are the cardinalities of these subsets of $\Omega$.
Here the M\"obius transform removes the overlaps between non-commuting observables.

For example,
\begin{eqnarray}
&&{\cal M}(\{r\})=\theta(\{r\});\;\;\;\;\;
{\cal M}(\{r,s\})=\theta (\{r,s\})-\theta(\{r\})-\theta(\{s\})\nonumber\\
&&{\cal M}(\{r,s,t\})=\theta (\{r,s,t\})-\theta(\{r,s\})-\theta(\{r,t\})-\theta(\{s,t\})+\theta(\{r\})+\theta(\{s\})+\theta(\{t\})
\end{eqnarray}

The inverse M\"obius transform is
\begin{eqnarray}\label{b7}
\theta (A)=\sum _{B\subseteq A}{\cal M}(B).
\end{eqnarray}
In this relation we put $A=\Omega$ and we get
\begin{eqnarray}
\sum _{B\subseteq \Omega}{\cal M}(B)={\bf 1}.
\end{eqnarray}
We rewrite this as 
\begin{eqnarray}\label{11AA}
\sum _r\theta(r)+\sum _{r,s}{\cal M}(\{r,s\})+\sum _{r,s,t}{\cal M}(\{r,s,t\})+...={\bf 1}.
\end{eqnarray}
This is a kind of resolution of the identity in terms of $\theta(r)$, but it also involves all their `higher' M\"obius transforms ${\cal M}(\{r,s\})$, ${\cal M}(\{r,s,t\})$, etc.
If additivity holds (if Eq.(\ref{66}) becomes equality), then all higher M\"obius transforms are zero (${\cal M}(\{r,s\})=0$, ${\cal M}(\{r,s,t\})=0$, etc).
This is the case if $n=d$ and $\theta(r)=\varpi_r$ (the set of orthonormal projectors in Eq.(\ref{C0})).
In this special case, and Eq.(\ref{11AA}) reduces to Eq.(\ref{C0}).

The M\"obius transform quantifies deviations from the additivity. 
Eq.(\ref{11AA}) involves the operators $\theta(r)$ and the higher M\"obius transforms which are corrections due to the fact that
additivity does not hold in general (Eq.(\ref{66}) is inequality).

\subsection{Approximate calculation of a density matrix from its expectation values}\label{sec16}

We consider $2^n$ $d\times d$ matrices $\theta(A)$ as in section \ref{sec14}.
If $R$ is an unknown  $d\times d$ density matrix, we assume that we measure the $n$ expectation values with the $n$ observables $\theta(\{1\}),...,\theta(\{n\})$:
\begin{eqnarray}\label{89}
\langle \theta(\{r\})\rangle={\rm Tr}[R\theta(\{r\})].
\end{eqnarray}
In the measurements we only use the $n$ observables corresponding to subsets with cardinality one.
The measurements need to be performed with different ensembles that describe the same density matrix, because these observables do not commute in general.

Below we introduce the corresponding quantum Choquet integral $C(R)$ and show that $R_1=\frac{{\cal C}(R)}{{\rm Tr}[{\cal C}(R)]}$ is an approximation to $R$, with error which 
we quantify with the set of errors $\{{\cal E}_1,...,{\cal E}_n\}$ for all observables, where
 \begin{eqnarray}\label{error}
{\cal E}_r=\left |\frac{{\rm Tr}[R_1\theta (\{r\})]-\langle \theta (\{r\})\rangle}{\langle \theta (\{r\})\rangle}\right |.
  \end{eqnarray}
But first we make the following to assumptions:
\begin{itemize}
\item
$d\le n<d^2-1$. 
Then we have incomplete information about $R$, and our aim is to find a family of {\bf positive semi-definite} matrices $R_1$ which have ${\rm Tr}[R_1\theta(\{r\})]$ approximately equal (with errors quantified in Eq.(\ref{error})) to the measured expectation values
$\langle \theta(\{r\})\rangle$.

Eq.(\ref{89}) has $n$ equations and $d^2-1$ real unknowns (related to the elements of the density matrix $R$).
We note that if we take $d^2-1-n$ of these unknowns to be free variables which take arbitrary values, and solve the system of $n$ equations with $n$ unknowns, we will get a matrix $R$ which in general is not 
positive semi-definite. 
In our approach the result is a positive semi-definite matrix, and the free variables are the matrices  $\theta(r,s), \theta(r,s,t),...$ which are not used in the measurements but enter in the calculation.

Apart from the practical value of this problem, there is theoretical merit in working with non-commuting operators and expectation values which are capacities (as opposed to the stronger concept of Kolmogorov probabilities which in a quantum context are related to orthonormal bases).
\item
In this section the $\theta(A)$ are positive definite, as opposed to positive semi-definite. This is because in the latter case there might be density matrices $R$ for
which all $\langle \theta(\{r\})\rangle=0$, and we want to avoid this. With this assumption all $\langle \theta(\{r\})\rangle>0$.
\item
Although not essential, we assume that the $\theta(\{1\}),...,\theta(\{n\})$  are linearly independent in the sense that
\begin{eqnarray}
a_1\theta(\{1\})+...+a_n\theta(\{n\})=0\Rightarrow a_1=...=a_n=0.
\end{eqnarray}
For linearly dependent observables, e.g, $\theta(\{3\})=\theta(\{1\})+\theta(\{2\})$,  measurements might give expectation values such that 
$\langle\theta(\{3\})\rangle\ne \langle\theta(\{1\})\rangle+\langle\theta(\{2\})\rangle$ due to experimental errors. Such contradictions in experimental data,  would increase the error 
in Eq.(\ref{error}).
In order to avoid such problems, we consider linearly independent observables.

We note that even in the case of linearly independent observables, we might have inequalities between the expectation values which the experimental values might violate due to experimental errors.
For example, if $A,B$ are positive semidefinite operators then ${\rm Tr}(AB)\le {\rm Tr}(A){\rm Tr}(B)$ (e.g., \cite{error}).
Therefore if $\theta(\{1\})\succ \theta(\{2\})$, then for any density matrix $R$ we get
\begin{eqnarray}
{\rm Tr}\{R[\theta(\{1\})-\theta(\{2\})]\}\le {\rm Tr}[\theta(\{1\})-{\rm Tr}[\theta(\{2\})]\implies \langle \theta(\{1\})\rangle -\langle \theta(\{2\}\rangle\le {\rm Tr}[\theta(\{1\})]-{\rm Tr}[\theta(\{2\})].
\end{eqnarray}
Due to experimental errors the expectation values $\langle \theta(\{1\})\rangle, \langle \theta(\{2\}\rangle$ might violate this, and then we will get large error in Eq.(\ref{error}).
\item
The error in Eq.(\ref{error}) depends on $n$. As $n$ increases we have more information about the system and we might expect the error to decrease. But this might not be the case, because parts of this information might contradict each other due to experimental errors (as discussed in the previous comment). 
\end{itemize}

We next consider a ranking permutation $\sigma$ of the elements of $\Omega$ (which depends on the density matrix), such that
\begin{eqnarray}\label{44}
0< \langle \theta [\{\sigma(1)\}\rangle ]\rangle \le \langle \theta [\{\sigma(2)\}]\rangle \le...\le \langle \theta [\{\sigma(n-1)\}]\rangle \le \langle \theta [\{\sigma(n)\}]\rangle\le 1.
\end{eqnarray}
Intuitively this means that the density matrix $R$ is very close to $\theta [\{\sigma(n)\}]$, less close to $\theta [\{\sigma(n-1)\}]$,
even less close to $\theta [\{\sigma(n-2)\}]$, etc.
Different density matrices are in general associated with different ranking permutations.
\begin{definition}
Density matrices which are associated with the same ranking permutation are called comonotonic. 
\end{definition}
Comonotonic density matrices are both very close to the same $\theta [\{\sigma(n)\}]$, less close to the same $\theta [\{\sigma(n-1)\}]$, etc,
and in this sense they have similar physical properties.

\subsection{Quantum Choquet integrals}\label{sec17}
Below we introduce Choquet integrals for matrices (not scalars) and we refer to them as quantum Choquet integrals.
\begin{proposition}\label{pro11}
We choose $2^n$ positive definite cumulative matrices $\theta(A)$ as in section \ref{sec14} (they obey Eq.(\ref{A})), and we assume that we know the $n$ expectation values $\langle \theta [\{r\}]\rangle={\rm Tr}[R\theta (\{r\})]$ where $R$ is an unknown density matrix. With a permutation $\sigma$ we reorder the expectation values as in Eq.(\ref{44}).

The quantum Choquet integral ${\cal C}(R)$  is the positive semi-definite matrix, given by one of the following expressions which are equivalent to each other:
\begin{itemize}
\item[(1)]
\begin{eqnarray}\label{00}
{\cal C}(R)&=&\Delta \langle \theta [\{\sigma (n)\}]\rangle \theta[\{\sigma (n)\}]+\Delta \langle \theta[\{\sigma (n-1)\}]\rangle \theta[\{\sigma (n-1),\sigma(n)\}]+...\nonumber\\&+&
\Delta \langle \theta [\{\sigma(n-d+1)\}]\rangle \theta[\{\sigma(n-d+1),...,\sigma(n)\}].
\end{eqnarray}
Here $\Delta\langle \theta[\sigma (n-k)]\rangle$ are the {\bf expectation value increments}:
\begin{eqnarray}\label{680}
k=0,...,d-2\;&\rightarrow&\;\Delta \langle \theta[\{\sigma (n-k)\}]\rangle=\langle \theta[\{\sigma (n-k)\}]\rangle-\langle \theta [\{\sigma (n-k-1)\}]\rangle\ge 0\nonumber\\
k=d-1\;&\rightarrow&\;\Delta \theta[\{\sigma(n-d+1)\}]=\langle\theta [\{\sigma(n-d+1)\}]\rangle\ge 0.
\end{eqnarray}
The $(n-d)$ smallest expectation values $\langle\theta[\{\sigma(1)\}]\rangle,..., \langle \theta[\{\sigma(n-d)\}]\rangle $ do not enter in the calculation of the Choquet integral ${\cal C}(R)$.
Only the $d$ largest expectation values enter in the calculation. 
Also the observables $\theta(r,s), \theta(r,s,t),...$ (whose  expectation vales have not been measured) do enter in the calculation.

\item[(2)]
\begin{eqnarray}\label{80}
{\cal C}(R)&=&\langle \theta [\{\sigma(n)\}]\rangle \Delta \theta[\{\sigma (n)\}]+\langle \theta[\{\sigma (n-1)\}]\rangle\Delta \theta[\{\sigma(n-1),\sigma(n)\}]+...\nonumber\\&+&\langle \theta[\{\sigma (n-d+1)\}\rangle\Delta \theta[\{\sigma(n-d+1),...,\sigma(n)\}]
\end{eqnarray}
Here $\Delta \theta[\{\sigma(n-k),...,\sigma(n)\}]$ are the {\bf increments of observables} (which are positive semi-definite matrices)
\begin{eqnarray}
k=1,...,d-1&\rightarrow&\Delta \theta[\{\sigma(n-k),...,\sigma(n)\}]=\theta[\{\sigma(n-k),...,\sigma(n)\}]-\theta[\{\sigma(n-k+1),...,\sigma(n)\}]\succ 0\nonumber\\
k=0&\rightarrow &\Delta\theta [\{\sigma(n)\}]=\theta[\{\sigma(n)\}]\succ 0.
\end{eqnarray}

\item[(3)]
The ${\cal C}(R)$ is given in terms of the M\"obius transforms in section \ref{sec15} as
\begin{eqnarray}
&&{\cal C}(R)={\cal C}_1(R)+{\cal C}_2(R)+...+{\cal C}_n(R)\nonumber\\
&&{\cal C}_1(R)=\sum _r\theta(\{\sigma(r)\})\langle \theta[\{\sigma(r)\}]\rangle \nonumber\\
&&{\cal C}_2(R)=\sum_{r,s}{\cal M}(\{\sigma(r),\sigma(s)\})\min_{r,s}(\langle \theta[\{\sigma(r)\}])\rangle,\langle\theta[\{\sigma(s)\}]\rangle )\nonumber\\
&&\vdots \nonumber\\
&&{\cal C}_n(R)={\cal M}(\{\sigma(1),...,\sigma(n)\})\langle \theta(\{\sigma(1)\})\rangle 
\end{eqnarray}
If additivity holds (if Eq.(\ref{66}) becomes equality) then ${\cal C}_2(R)=...={\cal C}_n(R)=0$.
This is the case if $n=d$ and $\theta(r)=\varpi_r$ (the set of orthonormal projectors in Eq.(\ref{C0})).
\end{itemize}

\end{proposition}
\begin{proof}
The proof of the equivalence of the two expressions is straightforward.

We prove the equivalence between the first expression and the third expression.
We replace all $\theta(A)$ in Eq(\ref{00}) with their M\"obius transform using Eq.(\ref{b7}).
Then the terms that contain the factor ${\cal M}(\{\sigma(r),\sigma(s)\})$ (where we assume $\langle \theta[\{\sigma(r)\}])\rangle\le \langle\theta[\{\sigma(s)\}]\rangle $) give
\begin{eqnarray}
\Delta \langle \theta[\{\sigma (r)\}]\rangle+...+\Delta \langle \theta [\{\sigma(n-d+1)\}]\rangle=\langle \theta[\{\sigma (r)\}]\rangle.
\end{eqnarray}
This gives the term ${\cal C}_2(R)$. In a similar way we get the other terms.
\end{proof}

It is easily seen that for comonotonic density matrices $R_1,R_2$, the Choquet integral corresponding to the density matrix $aR_1+(1-a)R_2$ is
\begin{eqnarray}
{\cal C}[aR_1+(1-a)R_2]=a{\cal C}(R_1)+(1-a){\cal C}(R_2);\;\;\;0\le a\le 1,
\end{eqnarray}
but in general this is {\bf not true}.
\begin{remark}
The calculation of Eq.(\ref{00}) involves multiplication of a number with a $d\times d$ matrix, which is repeated $d$ times.
We define the time complexity of a calculation, as the number of multiplications of complex numbers (this ignores additions).
Then the complexity of the calculation of the quantum Choquet integral ${\cal C}(R)$ is ${\cal O}(d^3)$.
In contrast the calculation of the classical Choquet integral ${\cal C}(f)$ in Eq.(\ref{480}) is ${\cal O}(d)$.
\end{remark}

\section{Examples of approximate calculation of a density matrix from its expectation values}
Given $n$ observables $\theta (\{1\}),...,\theta (\{n\})$ and their expectation values $\langle \theta (\{1\})\rangle,...,\langle \theta (\{n\})\rangle$
with an unknown density matrix $R$, we use the positive semidefinite matrix
\begin{eqnarray}
R_1=\frac{{\cal C}(R)}{{\rm Tr}[{\cal C}(R)]}
\end{eqnarray}
as an  approximation to the density matrix $R$.
As we already mentioned earlier the problem has more unknowns than the number of equations, and the $\theta(r,s), \theta(r,s,t),...$ are the analogue of the free variables in this approach. The result depends on the choice that we make for these matrices.

In section \ref{sec18} we consider the simple case of orthogonal projectors as observables, and we get a family of matrices with error in the expectation values close to zero (for $\lambda$ close to zero in Eq.(\ref{error})).
The interesting case is to have non-commuting observables and an example of this is in section \ref{sec19}.
We make two choices for the `free variables' and we get two density matrices.

An interesting open question for given observables $\theta (\{1\}),...,\theta (\{n\}$, is to find the set of $n$-tuples 
$(\langle \theta (\{1\})\rangle,...,\langle \theta (\{n\})\rangle)$ for all density matrices:
\begin{eqnarray}\label{45}
{\cal S}[\theta (\{1\}),...,\theta (\{n\})]=\{(\langle \theta (\{1\})\rangle,...,\langle \theta (\{n\})\rangle)\}\subset {\mathbb R}^n.
\end{eqnarray}

 \subsection{Example with orthogonal projectors as observables}\label{sec18}
 Let $\ket{v_r}$ be an orthonormal basis in the Hilbert space $H(3)$ and $e_r$ numbers with 
 \begin{eqnarray}
 0<e_1<e_2<e_3;\;\;\;e_1+e_2+e_3=1.
 \end{eqnarray}
 We choose $n=d=3$ and $\theta(\{r\})=\pi(r)=\ket{v_r}\bra{v_r}$ and we assume that an unknown density matrix $R$ gives the expectation values $\langle\theta(r)\rangle=\bra{v_r}R\ket{v_r}=e_r$.
We will find a family of density matrices with expectation values approximately $e_r$ (in addition to the trivial solution  $\sum e_r\pi_r$).

In this example Eq.(\ref{680}) gives
\begin{eqnarray}
\Delta \langle \theta \{1\}\rangle =e_1;\;\;\;\Delta \langle \theta \{2\}\rangle =e_2-e_1;\;\;\;\Delta \langle \theta \{3\}\rangle =e_3-e_2.
\end{eqnarray}
For the cumulative matrices we choose  diagonal matrices (in the basis $\{\ket{v_r}\}$) with diagonal elements :
\begin{eqnarray}
\theta(\{1,2\})={\rm diag}(1\;\;1\;\;0);\;\;\;\theta(\{1,3\})={\rm diag}(1\;\;0\;\;1);\;\;\;\theta(\{2,3\})={\rm diag}(\lambda\;\;1\;\;1);\;\;\;\theta(\{1,2,3\})={\bf 1}
\end{eqnarray}
Here $0\le \lambda<1$. The requirements in Eq(\ref{A}) are satisfied.

In this case we find
\begin{eqnarray}\label{82}
{\cal C}(R)={\rm diag}((1-\lambda)e_1+\lambda e_2\;\;e_2\;\;e_3);\;\;\;{\rm Tr}[{\cal C}(R)]=1+\lambda(e_2-e_1).
\end{eqnarray}
From Eq.(\ref{82}) follows that:
\begin{eqnarray}
R_1=\frac{{\cal C}(R)}{{\rm Tr}[{\cal C}(R)]}=\frac{1}{1+\lambda (e_2-e_1)}{\rm diag}((1-\lambda)e_1+\lambda e_2\;\;e_2\;\;e_3).
\end{eqnarray}
The error in the expectation values as in Eq.(\ref{error}), is
\begin{eqnarray}
{\cal E}_1=\frac{\lambda (e_2-e_1)(1-e_1)}{e_1[1+(e_2-e_1)\lambda]};\;\;\;{\cal E}_2=\frac{\lambda (e_2-e_1)}{1+(e_2-e_2)\lambda};\;\;\;
{\cal E}_3=\frac{\lambda (e_2-e_1)}{1+(e_2-e_1)\lambda}.
\end{eqnarray}
Now we found a family of density matrices (for various values of $\lambda$) with expectation values close to the measured ones $\langle\theta(r)\rangle=e_r$.
The result depends on $\lambda$, and for $\lambda$ close to zero the error is close to zero.
If $\lambda=0$ we get the trivial solution ${\cal C}(R)=\sum e_r\pi_r$.

\subsection{Example with non-commuting observables}\label{sec19}
In this example $d=3$ and $n=4$.
We make two choices for the operators $\theta(A)$, which are shown in table \ref{T1}, and satisfy Eq.(\ref{A}).
The $\theta(1), \theta(2), \theta(3), \theta(4)$ are the same in both cases, but some of the rest $\theta(A)$ are different.
We emphasise that in the present approach the result is always a positive semidefinite matrix.

The chosen expectation values $\langle \theta (\{1\})\rangle$,...,$\langle \theta (\{4\})\rangle)$, need to belong to the set ${\cal S}[\theta (\{1\}),...,\theta (\{4\})]$ in Eq.(\ref{45}).
In order to ensure that  we start with the density matrix
\begin{eqnarray}
R=\frac{1}{10}\begin{pmatrix}
3&3&0\\
3&4&i\\
0&-i&3
\end{pmatrix}
\end{eqnarray}
which has the eigenvalues $0.030$, $0.300$, $0.670$, and in an ideal experiment will give the expectation values
 \begin{eqnarray}\label{48}
\langle \theta (\{1\})\rangle=0.125;\;\;\;\;\langle \theta (\{2\})\rangle=0.200;\;\;\;\;\langle \theta (\{3\})\rangle=0.155;\;\;\;\;\langle \theta (\{4\})\rangle=0.170.
  \end{eqnarray}
  Of course there are many density matrices that give the same expectation values.
  Our aim is to find density matrices that give approximately these expectation values.
  
The ordering in this case is
 \begin{eqnarray}
\sigma(1)=1;\;\;\;\sigma(2)=3;\;\;\;\sigma(3)=4;\;\;\;\sigma(4)=2.
\end{eqnarray}
Then
\begin{eqnarray}
{\cal C}(R)&=&\Delta \langle \theta [\{\sigma (4)\}]\rangle \theta[\{\sigma (4)\}]+\Delta \langle \theta[\{\sigma (3)\}]\rangle \theta[\{\sigma (3),\sigma(4)\}]+
\Delta \langle \theta [\{\sigma(2)\}]\rangle \theta[\{\sigma(2),\sigma(3),\sigma(4)\}].
\end{eqnarray}
From the above expectation values we get
\begin{eqnarray}
\Delta \langle \theta [\{\sigma (4)\}]\rangle=0.030;\;\;\;\Delta \langle \theta [\{\sigma (3)\}]\rangle=0.015;\;\;\;\Delta \langle \theta [\{\sigma (2)\}]\rangle=0.030.
\end{eqnarray}
Therefore
\begin{eqnarray}\label{W}
{\cal C}(R)&=&0.030 \theta(\{2\})+0.015\theta(\{2,4\})+0.030\theta(\{2,3,4\}).
\end{eqnarray}

For the first choice of $\theta(A)$ this becomes
\begin{eqnarray}
{\cal C}(R)&=&0.075 \theta(\{2\})+0.036\theta(\{3\})+0.045\theta(4\}).
\end{eqnarray}
In this case we get the density matrix
\begin{eqnarray}
R_1=\frac{{\cal C}(R)}{{\rm Tr}[{\cal C}(R)]}=\begin{pmatrix}
0.545&   0.131 + 0.025i&   0.104 + 0.063i\\
   0.131 - 0.025i&   0.205 &   0.012\\
   0.104 - 0.063i&   0.012&  0.248\\
\end{pmatrix},
\end{eqnarray}
which has eigenvalues $0.147$, $0.223$, $0.630$, and expectation values
 \begin{eqnarray}\label{W1}
{\rm Tr}[R_1\theta (\{1\})]=0.140;\;\;\;{\rm Tr}[R_1\theta (\{2\})]=0.220;\;\;\;{\rm Tr}[R_1\theta (\{3\})]=0.220;\;\;\;{\rm Tr}[R_1\theta (\{4\})]=0.190
  \end{eqnarray}

For the second choice of $\theta(A)$ we get 
\begin{eqnarray}
{\cal C}(R)&=&0.075 \theta(\{2\})+0.036\theta(\{3\})+0.054\theta(4\}).
\end{eqnarray}
This leads to the density matrix
\begin{eqnarray}
R_2=\frac{{\cal C}(R)}{{\rm Tr}[{\cal C}(R)]}=\begin{pmatrix}
0.536 &   0.123 + 0.028i&   0.097 + 0.071i\\
   0.123 - 0.028i&   0.217 &  0.014\\
   0.097 - 0.071i&   0.014&  0.245\end{pmatrix},
\end{eqnarray}
which has eigenvalues $0.158$, $0.222$, $0.619$, and expectation values
 \begin{eqnarray}\label{W2}
{\rm Tr}[R_2\theta (\{1\})]=0.138;\;\;\;{\rm Tr}[R_2\theta (\{2\})]=0.216;\;\;\;{\rm Tr}[R_2\theta (\{3\})]=0.216;\;\;\;{\rm Tr}[R_2\theta (\{4\})]=0.192.
  \end{eqnarray}
We calculate the error in the various observables with Eq.(\ref{error}). For Eq.(\ref{W1}) we find 
\begin{eqnarray}
{\cal E}_1=0.104;\;\;\;{\cal E}_2=0.080;\;\;\;{\cal E}_3=0.393;\;\;\;{\cal E}_4=0.129.
  \end{eqnarray}
For Eq.(\ref{W2}) we find 
\begin{eqnarray}
{\cal E}_1=0.120;\;\;\;{\cal E}_2=0.100;\;\;\;{\cal E}_3=0.419;\;\;\;{\cal E}_4=0.117.
  \end{eqnarray}
The error  is different in the various observables, and is the price that we pay for finding positive semidefinite matrices.

\section{Discussion and open problems}

We studied the set $M$ of Hermitian matrices as a partially ordered set with the  L\"{o}wner partial order.
Several properties of this partial order have been presented in section 3 (propositions \ref{proA1}, \ref{proA2},\ref{L0}).
We have also shown (section \ref{sec100}) that many quantities used in other methodologies are non-comparable, and this differentiates our approach which is based on cumulative (comparable) quantities from them.

Upper and lower sets in $M$ define the concept of cumulativeness  in the context of Hermitian matrices. Their properties  have been discussed in section \ref{sec101}, 
and several examples have been given in section \ref{sec29}.

The topological approach in section 5, complements the partial order approach.
The  set $M$ of Hermitian matrices is studied as a topological space with the upper and lower sets, playing the role of open and closed sets correspondingly.
We have shown that the set $M$ of Hermitian matrices is a $T_0$ topological space, and interpreted this physically (section \ref{secA5}).
We have also shown that its subset ${\mathfrak D}$ of density matrices, is Hausdorff totally disconnected topological space (section \ref{secA6}).

The study of the set of Hermitian matrices as a partially ordered set and also as a topological space, is a `toolbox' interesting in its own right.
But here it has been  used within a formalism for capacities, cumulative functions and quantum Choquet integrals with Hermitian matrices. 
This formalism requires an ordering (e.g. Eq.(\ref{A})) which is trivial for scalar quantities, but here we have matrices and it requires careful definition (as in sections \ref{sec37}, \ref{sec101}, \ref{topo}).
In remark \ref{rem18} we gave an example of using the partial order theory language, and the topological language in the context of capacities and quantum Choquet integrals. 

We have used the Choquet formalism (proposition \ref{pro11}) to calculate an unknown density matrix with given expectation values (with respect to $n$ observables).
This problem can be viewed as a generalisation of quantum tomography, which uses very specific observables. In the present work we use arbitrary observables, and the approach is very different.
We have shown that $\frac{{\cal C}(R)}{{\rm Tr}[{\cal C}(R)]}$ is a density matrix with approximately these expectation values, and with errors which have been quantified. We gave several examples (in sections \ref{sec18}, \ref{sec19}).
In addition to the practical value of this problem, there is the theoretical value of approaching non-commutativity in a quantum context, with capacities and quantum Choquet integrals.

Extension of the present work can be in various directions as follows:
\begin{itemize}
\item
The use of Kolmogorov probabilities in a quantum context leads to violations of inequalities.
The use of capacities (which replace Eqs.(\ref{1Z}), (\ref{B}) for Kolmogorov probabilities with the weaker relation in Eq(\ref{A2})) offers a different approach to the nature of quantum probabilities.
In a classical context, capacities are used widely in areas like Artificial Intelligence, Big Data  and Mathematical Economics, in problems where  the various alternatives are not
truly independent.  The analogue of truly independent alternatives in a quantum context, is the use of commuting observables related to orthonormal bases. For non-commuting observables (as in section \ref{sec19}), the use of capacities, cumulative quantities, M\"obius transforms and quantum Choquet integrals, offers a new approach to 
non-commutativity.
\item
The study of the set ${\cal S}[\theta (\{1\}),...,\theta (\{n\})]$ in Eq.(\ref{45}), as a subset of ${\mathbb R}^n$, for non-commuting observables $\theta (\{1\}),...,\theta (\{n\})$, is an open problem.
Again this looks at non-commutativity from a different angle.
\item
In this paper we studied partial order and topology of Hermitian matrices, and
this is appropriate for  observables. But it will be interesting to extend this to non-Hermitian matrices. 
The fact that their eigenvalues are complex numbers, makes this problem challenging.
\end{itemize}
The work brings together ideas from partial order theory, topology, and Choquet integrals and offers a new approach to quantum non-commutativity.
It is a contribution to the developing area that combines quantum techniques with Artificial Intelligence.

\newpage
\begin{table}[ht]
\caption{The $\theta(A)$ for the example in section \ref{sec18}. They satisfy Eq.(\ref{A}).}
{\begin{tabular}{|c|c|c|}\\
&{\rm first choice}&{\rm second choice}\\\hline
$\theta(\emptyset)$&${\bf 0}$&${\bf 0}$\\\hline
 $\theta (\{1\})$&$\begin{pmatrix}
 0.15 &0.05& 0.05\\
    0.05 &0.05& 0\\
    0.05 &0& 0.1
    \end{pmatrix}$&$\begin{pmatrix}
 0.15 &0.05& 0.05\\
    0.05 &0.05& 0\\
    0.05 &0& 0.1
    \end{pmatrix}$\\\hline
$\theta (\{2\})$&$\begin{pmatrix}
0.25& 0.1 &0.05\\
    0.1& 0.05& 0\\
    0.05& 0& 0.15
\end{pmatrix}$&$\begin{pmatrix}
0.25& 0.1 &0.05\\
    0.1& 0.05& 0\\
    0.05& 0& 0.15
\end{pmatrix}$\\\hline
$\theta (\{3\})$&$\begin{pmatrix}
0.3 &0.05 &0.1\\
    0.05 &0.05&0\\
    0.1&0& 0.05
\end{pmatrix}$&$\begin{pmatrix}
0.3 &0.05 &0.1\\
    0.05 &0.05&0\\
    0.1&0& 0.05
\end{pmatrix}$\\\hline
$\theta (\{4\})$&$\begin{pmatrix}
0.2 &0.04i &0.1i\\
    -0.04i& 0.02& 0.02\\
    -0.1i &0.02 &0.1
\end{pmatrix}$&$\begin{pmatrix}
0.2 &0.04i &0.1i\\
    -0.04i& 0.02& 0.02\\
    -0.1i &0.02 &0.1
\end{pmatrix}$\\\hline
$\theta (\{1,2\})$&$\theta (\{1\})+1.3\theta (\{2\})$&$\theta (\{1\})+1.3\theta (\{2\})$\\\hline
$\theta (\{1,3\})$&$\theta (\{1\})+\theta (\{3\})$&$\theta (\{1\})+\theta (\{3\})$\\\hline
$\theta (\{1,4\})$&$\theta (\{1\})+\theta (\{4\})$&$\theta (\{1\})+\theta (\{4\})$\\\hline
$\theta (\{2,3\})$&$\theta (\{2\})+1.2\theta (\{3\})$&$\theta (\{2\})+1.2\theta (\{3\})$\\\hline
$\theta (\{2,4\})$&$\theta (\{2\})+\theta (\{4\})$&$\theta (\{2\})+1.2\theta (\{4\})$\\\hline
$\theta (\{3,4\})$&$\theta (\{3\})+\theta (\{4\})$&$\theta (\{3\})+\theta (\{4\})$\\\hline
$\theta (\{1,2,3\})$&$\theta (\{1\})+1.3\theta (\{2\})+1.2\theta (\{3\})$&$\theta (\{1\})+1.3\theta (\{2\})+1.2\theta (\{3\})$\\\hline
$\theta (\{1,2,4\})$&$\theta (\{1\})+1.3\theta (\{2\})+\theta (\{4\})$&$\theta (\{1\})+1.3\theta (\{2\})+1.2\theta (\{4\})$\\\hline
$\theta (\{1,3,4\})$&$\theta (\{1\})+\theta (\{3\})+\theta (\{4\})$&$\theta (\{1\})+\theta (\{3\})+\theta (\{4\})$\\\hline
$\theta (\{2,3,4\})$&$\theta (\{2\})+1.2\theta (\{3\})+\theta (\{4\})$&$\theta (\{2\})+1.2\theta (\{3\})+1.2\theta (\{4\})$\\\hline
$\theta (\{1,2,3,4\})$&${\bf 1}$&${\bf 1}$\\\hline
\end{tabular}}
\label{T1}
\end{table}

\end{document}